\documentclass[intlimits,twoside,a4paper]{article}

\usepackage{amsmath,amssymb}
\usepackage{graphicx}
\usepackage{wrapfig}
\usepackage[T2A]{fontenc}
\usepackage[cp1251]{inputenc}

\usepackage{amsfonts}
\usepackage{psfrag}
\usepackage{color}
\usepackage{cancel}
\usepackage{ulem}

\usepackage[eqsecnum]{cmpj2}

\issue{2012}{15}{2}{23605}
\doinumber{10.5488/CMP.15.23605}

\newcommand{\be}{\begin{equation}}
\newcommand{\ee}{\end{equation}}
\newcommand{\bea}{\begin{eqnarray}}
\newcommand{\eea}{\end{eqnarray}}
\input{epsf}

\title[Polydisperse dipolar hard-sphere fluid]%
{Liquid-gas phase behavior of polydisperse dipolar
hard-sphere fluid: Extended thermodynamic perturbation theory for central
force associating potential}
\author[Yu.V. Kalyuzhnyi, S. Hlushak, P.T. Cummings]{Yu.V. Kalyuzhnyi\refaddr{label1},
        S. Hlushak\refaddr{label1,label2}, P.T. Cummings\refaddr{label2,label3}}
\addresses{
\addr{label1} Institute for Condensed Matter Physics of the National
Academy of Sciences of Ukraine, \\ 1 Svientsitskii Str., 79011 Lviv, Ukraine
\addr{label2} Department of Chemical Engineering, Vanderbilt University,
Nashville, Tennessee 37235
\addr{label3} Nanomaterials Theory Institute,
Center for Nanophase Material Sciences, Oak Ridge National
Laboratory, \\ Oak Ridge, Tennessee 37830
}

\authorcopyright{Yu.V. Kalyuzhnyi, S. Hlushak, P.T. Cummings, 2012}
\date{Received January 31, 2012, in final form April 5, 2012}

\begin{document}

\maketitle

\begin{abstract}

The liquid-gas phase diagram for polydisperse dipolar hard-sphere fluid with polydispersity in the hard-sphere size and dipolar moment is calculated
using extension of the recently proposed thermodynamic perturbation theory for central
force (TPT-CF) associating potential. To establish the connection with the phase behavior of ferrocolloidal dispersions it is assumed that the dipole moment is proportional to
the cube of the hard-sphere diameter. We present and discuss the full phase diagram, which includes cloud and shadow curves, binodals and distribution functions of the coexisting daughter phases at different degrees of the system polydispersity.
In all cases studied polydispersity increases the region of the phase instability and shifts the critical point to the higher values of the temperature and density.
The larger size particles always fractionate to the liquid phase and the smaller size
particles tend to move to the gas phase. At relatively high values
of the system polydispersity three-phase coexistence is observed.
\pacs
64.10.+h, 64.70.Fx, 82.70.Dd

\end{abstract}

\section{Introduction}

\looseness=-1In this paper we consider the liquid-gas phase behavior of polydisperse dipolar hard-sphere
mixture. Recently, liquid-gas phase equilibria in monodisperse dipolar hard-sphere fluid, Yukawa
dipolar hard-sphere fluid and Shtockmayer fluid were studied using thermodynamic perturbation theory for
central force (TPT-CF) associating potential~\cite{kalstell,dipole,dipoleYuk,shtock}.
In this study we propose an extension of the TPT-CF, which enables us to investigate the phase behavior of
polydisperse mixture of the dipolar hard spheres with polydispersity in both hard-sphere size
and dipole moment. We call this extension as extended TPT-CF (ETPT-CF).
Similar to our previous study~\cite{RTPT_multi}, ETPT-CF combines Wertheim's TPT~\cite{w12,w34}
for associating fluid with association due to off-center attractive sites, and TPT-CF~\cite{dipole,RTPT_one},
which permits a multiple bonding of one site. In our theory we have several Wertheim's
types of associating sites with the possibility for each site to be multiply bonded
(in Wertheim's TPT each site is only singly bondable). Final expressions for thermodynamical
properties of polydisperse dipolar hard-sphere fluid is written in terms of the finite number
of distribution function moments, i.e., in the framework of ETPT-CF this system belongs to the
family of the so-called truncatable free energy models (see~\cite{sollich} and references
therein). This property enables us to calculate
the full liquid-gas phase diagram (including cloud and shadow curves and binodals) and to
study the effects of fractionation on the level of the distribution functions of coexisting daughter~phases.

\section{Extended thermodynamic perturbation theory for central force  associative potential}

\subsection{Analysis and classification of diagrams}

We consider a multicomponent fluid mixture consisting of $n$ species with a number density
$\rho=\sum_a^n\rho_a$ at a temperature $T$ ($\beta=1/k_\mathrm{B}T$), where $\rho_a$ is the density of
the particles of $a$ species. The particles of the species $a$ and $b$ interact via the
pair potential $U_{ab}(12)$, which can be written as a sum of the reference
$U_\mathrm{ref}^{ab}(12)$ and associative $U_\mathrm{ass}^{ab}(12)$ parts
\be
U_{ab}(12)=U^{ab}_\mathrm{ref}(12)+U_\mathrm{ass}^{ab}(12),
\label{utot}
\ee
where $1$ and $2$ denote positions and orientations of the particles $1$ and $2$.
We assume that the associative part of the potential can be represented as a sum of
$M_a\times M_b$ terms, i.e.
\be
U^{ab}_\mathrm{ass}(12)=\sum_{KL}U_{KL}^{ab}(12),
\label{uass}
\ee
where the lower indices $K$ and $L$ take the values $\underbrace{A,B,C,\ldots}_{M^{_a}}$
and $\underbrace{A,B,C,\ldots}_{M_{b}}$, respectively. These values specify the splitting
of the total associating potential $U_\mathrm{ass}^{ab}(12)$ into several particular pieces.
For example in the case of the models utilized by Wertheim~\cite{w34} these indices denote
off-center attractive sites and in the case of the Mercedes-Benz (MB)
type of models~\cite{MB} or cone models~\cite{chapman} they stand for the type of hydrogen bonding
arms. Hereafter we will refer to these indices as the site indices, keeping in mind
that they may have a more general meaning. Here $M_a$ and $M_b$ are the number of such sites
on the particles of $a$ and $b$ species, respectively. According to (\ref{utot}) and
(\ref{uass}) the Mayer function $f_{ab}(12)$ for the total potential~(\ref{utot}) takes the
following form:
\be
f_{ab}(12)=f_\mathrm{ref}^{ab}(12)+e_\mathrm{ref}^{ab}(12)
\left\{\prod_{KL}\left[1+f^{ab}_{KL}(12)\right]-1\right\},
\label{fsplit}
\ee
where we use the usual notation:
\be
e(12)=\exp{\left[-\beta U(12)\right]},
\qquad
f(12)=e(12)-1.
\label{ef}
\ee

For the sake of diagrammatic analysis we will follow Wertheim~\cite{w34} and instead of
circles we introduce hypercircles to represent particles in diagrammatic expansions.
Each hypercircle is depicted as a large open circle with small circles inside denoting
the sites. Corresponding cluster integrals are represented by the diagrams built on a
hypercircles connected by $f_\mathrm{ref}$ and $e_\mathrm{ref}$ bonds
and site circles connected by
the associating bonds $f_{KL}$.
Due to the decomposition of the Mayer function $f_{ab}(12)$
(\ref{fsplit}) we will have the following diagrammatic expressions for the logarithm of a
grand partition function $\Xi$ and for the one-point density $\rho_a(1)$ in terms of the
activity $z$:
\\\\
{\it
$\ln \Xi=$ sum of all topologically distinct connected diagrams consisting of field
${\tilde z}$ hypercircles, $f_\mathrm{ref}$, $e_\mathrm{ref}$ and $f_{KL}$ bonds. Each bonded pair of
${\tilde z}$ hypercircles has either $f_\mathrm{ref}$, or $e_\mathrm{ref}$ and one or more $f_{KL}$
bonds.}
\\\\
{\it
$\rho_a(1)=$ sum of all topologically distinct connected diagrams obtained from
$\ln \Xi$ by replacing in all possible ways one field ${\tilde z}$ hypercircle by a
${\tilde z}_a(1)$ circle labeled $1$}.
\\\\
Here ${\tilde z}_a(i)=z_a\exp\left[-\beta U_a(i)\right]$, $i$ denotes position and orientation of
the particle $i$, and $U_a(i)$ is an external field. For a uniform system
${\tilde z}_a(1)\equiv z_a$.
Following~\cite{w12,w34,kalstell,dipole} we introduce the
definition of the $s$-mer diagrams.
These are the diagrams consisting of $s$ hypercircles, which all are connected
by the network of
$f_{KL}$ bonds.
The site circles, which are incident with more than $m^a_K$ $f^{ab}_{KL}$ bonds are called
{\it oversaturated} site circles.
We consider now the set of oversaturated site circles with each pair connected by at least
one path formed by the circles from the same set. The subdiagram involving this set of
circles, together with the site circles adjacent to them and $f_{KL}$ bonds connecting all
these circles , we call the {\it oversaturated} subdiagram.
The set of all possible $s$-mer diagrams can be constructed
in three steps: (i) generating the subset of all possible connected diagrams
with only $f_{KL}$ bonds, (ii) inserting combined bond $e_\mathrm{ref}=f_\mathrm{ref}+1$
between all pairs of hypercircles with the site circles, which belong to the same maximal
oversaturated subdiagram and (iii) taking all ways of
inserting an $f_\mathrm{ref}$ bond between the pairs of hypercircles,
which were not connected during the previous two steps. As a result the diagrams, which
appear in $\ln \Xi$ and $\rho(1)$, can be expressed in terms of the $s$-mer diagrams:
\\\\
{\it
$\ln\Xi$ = sum of all topologically distinct connected diagrams consisting of $s$-mer
diagrams with $s=1,\ldots,\infty$ and $f_\mathrm{ref}$ bonds between pairs of hypercircles in
distinct $s$-mer diagrams.}
\\\\
The procedure for obtaining the expression for $\rho_a(1)$ from $\ln\Xi$ remains unchanged.

The diagrams appearing in the ${\tilde z}$ expansion of the singlet density $\rho_a(1)$
can be classified with respect to the number of $f^{ab}_{KL}$ bonds associated with the
labeled ${\tilde z}_a(1)$ hypercircle. We denote the sum of the diagrams with
$i_K\leqslant m^a_K$ associating bonds connected to the site $K$ ($K=A,B,C,\ldots$),
which belongs to the particle of species $a$ as
$\rho^a_{A_{i_A},B_{i_B},C_{i_C},\ldots}(1)$. Any site K, which is connected to $i_K > m^a_K$
associating bons, will be denoted as $K_{m^a_K}$
In what follows we will use also a condensed version of the
notation, i.e.
\be
\rho^a_{A_{i_A},B_{i_B},C_{i_C},\ldots}(1)\equiv \rho^a_{i_A,i_B,i_C,\ldots}(1)\equiv
\rho^a_{\left\{i\right\}}(1),
\label{state}
\ee
where $\left\{i\right\}=i_A,i_B,i_C,\ldots$.
The set $\left\{i\right\}$ with all indices, except one index $i_K$, equal 0, will be
denoted as $i_K$, i.e. $\left\{i\right\}=0,\dots,0,i_K,0,\ldots,0\equiv i_k$, so that
for any quantity $x^a_{\left\{i\right\}}$ we have
\be
x^a_{\left\{i\right\}}=x^a_{0,\dots,0,i_K,0,\ldots,0}\equiv x^a_{K_{i_K}}\equiv x^a_{i_K}\, .
\label{example1}
\ee
Thus $\rho_a(1)$ can be written as follows
\be
\rho_a(1)=\sum_{i_A,i_B,\ldots=0}^{m^a_A,m^a_B,\ldots}
\rho^a_{A_{i_A},B_{i_B},C_{i_C},\ldots}(1)\equiv
\sum^{\left\{m^a\right\}}_{\left\{i\right\}=0}\rho^a_{\left\{i\right\}}(1).
\label{rhotot}
\ee

\subsection{Topological reduction}

The process of switching from the activity to a density
expansion goes in the same fashion as in Refs.~\cite{w34,kalstell,dipole}.
However, to proceed it is convenient to use an operator form of notation. The operators
are introduced in a manner similar to that presented in references~\cite{w34,kalstell}
to which we refer the reader for more details. We associate with each labeled $l$
hypercircle an operator $\epsilon^a_{\left\{i\right\}}(l)$ with the following
properties:
\be
\begin{array}{ll}
&\epsilon^a_{\left\{i\right\}}(l)=0, \qquad \text{ if any} \qquad i_K > m_K^a\,,\\
&\epsilon^a_{\left\{i\right\}}(l)=1, \qquad \text{ if all} \qquad \ \ i_K=0\,,\\
&\epsilon^a_{\left\{i\right\}}(l)\epsilon^a_{\left\{j\right\}}(l)=
\epsilon^a_{\left\{i+j\right\}}(l)\,,
\end{array}
\label{oper}
\ee
where $\left\{i+j\right\}\equiv i_A+j_A,i_B+j_B,i_C+j_C,\ldots$.
The one-point quantities, which, for convenience, are denoted by $x^a_{\left\{i\right\}}$,
can be presented as illustrated below:
\be
{\hat x}_a(1)=\sum_{\left\{i\right\}=0}^{\left\{m^a\right\}}
\epsilon^a_{\left\{i\right\}}(1)x^a_{\left\{i\right\}}(1).
\label{hat}
\ee
The operators $\epsilon^a_{\left\{i\right\}}$ are straightforward generalization of
the operators introduced earlier~\cite{w34,kalstell,kalvojko}. Thus, the rules of
manipulation with the new quantities ${\hat x}_a$ are similar to that discussed before.
In particular, the usual algebraic rules apply to these quantities and analytical
functions of ${\hat x}_a$ are defined by the corresponding power series.
Similar, as in references~\cite{w34,kalstell,kalvojko}, it is convenient to use the
angular brackets to specify the operation
\be
\langle {\hat x}_a\rangle=x^a_{\left\{m^a\right\}}\,.
\label{brack}
\ee
In the case of several labeled circles the subscripts on the brackets denote the circle
to which procedure (\ref{brack}) is to be applied.

Analyzing the connectivity of the diagrams in $\rho_a(1)$, at a
labeled ${\tilde z}_a(1)$ hypercircle we have
\be
{\hat \rho}_a(1)/{\tilde z}_a(1)=\exp{\left[{\hat c}_a(1)\right]},
\label{rhoexp}
\ee
where $c^a_{\left\{i\right\}}(1)$ with $\left\{i\right\}\ne\left\{0\right\}$
denotes the sum of diagrams in
$\rho^a_{\left\{i\right\}}(1)/\rho^a_{\left\{0\right\}}(1)$
for which the labeled 1 hypercircle is not an articulation circle.
Similarly $c^a_{\left\{0\right\}}(1)$ denotes the sum of diagrams in
$\rho^a_{\left\{0\right\}}(1)/{\hat z}_a(1)$ for which
hypercircle 1 is not an articulation circle.
Elimination of the diagrams containing field articulation circles can be achieved by
switching from an activity to a density expansion.
To do so we adopt the following
rule: each field hypercircle ${\tilde z}_a$, with bonding state of its sites represented
by the set $\left\{l\right\}$, in all irreducible diagrams
${\hat c}_a$ is replaced by a $\sigma^a_{\left\{j\right\}}$
hypercircle, where
$j_K=m^a_K-l_K$ ($K=A,B,\ldots$) for $m^a_K-l_K \geqslant 0$ and $j_K=0$ for
$m^a_K-l_K < 0$. The new quantities $\sigma^a_{\left\{i\right\}}(1)$ are connected
to the densities $\rho^a_{\left\{i\right\}}(1)$ via the following relation:
\be
{\hat \sigma}_a(1)={\hat \rho}_a(1)\sum_{\left\{i\right\}=0}^{\left\{m^a\right\}}
\epsilon^a_{\left\{i\right\}}(1).
\label{sigma}
\ee
This relation can be inverted expanding $\left[\sum_{\left\{i\right\}=0}^{\left\{m^a\right\}}
\epsilon^a_{\left\{i\right\}}(1)\right]^{-1}$ into a power series, i.e.
\be
{\hat \rho}_a(1)={\hat \sigma}_a(1)\prod_{K=A}\left[1-\epsilon^a_{K_1}(1)
\right].
\label{inv}
\ee

Now the diagrammatic expansions for $c^a_{\left\{i\right\}}$ can be expressed in terms
of the irreducible diagrams. To present this result in compact and convenient form we introduce
a sum of the diagrams $c^{(0)}$ defined as follows:
\\\\
{\it
$c^{(0)}$ = sum of all topologically distinct irreducible diagrams consisting of $s$-mer
diagrams with $s=1,\ldots,\infty$ and $f_\mathrm{ref}$ bonds between pairs of hypercircles in distinct $s$-mer
diagrams. All hypercircles are field circles carrying the $\sigma$-factor according to the rule
formulated above}
\\\\
Functional differentiation of $c^{(0)}$ with respect to $\sigma^a_{\left\{m^a-i\right\}}$ gives an
expression for $c^a_{\left\{i\right\}}$:
\be
c^a_{\left\{i\right\}}=\frac{\delta c^{(0)}} {\delta \sigma^a_{\left\{m^a-i\right\}}}\,.
\label{fder}
\ee

\subsection{Extended thermodynamic perturbation theory for central force associating \\ potential}

Now we are in a position to rewrite the regular one-density virial expansion for the pressure
$P$ in terms of the density parameters ${\hat \sigma}_a(1)$. Following the scheme,
proposed earlier~\cite{w12,w34,dipole,shtock} we have expression for the pressure in operator
form
\be
\beta PV=\sum_a\int \langle{\hat \sigma}_a(1)\left[1-{\hat c}_a(1)\right]
\rangle\;\rd\left(1\right)+c^{(0)}
\label{operPV}
\ee
and explicitly
\be
\beta PV=\sum_a\int\left[\rho_a(1)-\sum_{\left\{i\right\}=0}^{\left\{m^a\right\}}
\sigma^a_{\left\{m^a-i\right\}}(1)c^a_{\left\{i\right\}}(1)\right]\;\rd(1)+c^{(0)},
\label{explicitPV}
\ee
where $V$ is the volume of the system.
Similarly, as in~\cite{w12,w34} one can verify that these expressions satisfy the regular
thermodynamic relation
$
{\bar \rho}_a=\beta\partial P/ \partial \mu_a,
$
where ${\bar \rho}_a=\int\rho_a(1)\;\rd(1)$ and $\mu^a$ is the chemical potential.
This can be achieved by taking a variation of (\ref{operPV}) (or (\ref{explicitPV}))
 and combining~(\ref{rhoexp}),
(\ref{inv}) and (\ref{fder}).
The corresponding expression for Helmholtz free energy is
\be
\beta A=\sum_a\int\left[\rho_a(1)\ln{\frac{\rho^a_{\left\{0\right\}}(1)}{\Lambda_a}}+
\sum_{\left\{i\right\}\ne 0}^{\left\{m^a\right\}}
\sigma^a_{\left\{m^a-i\right\}}(1)c^a_{\left\{i\right\}}(1)\right]\;\rd(1)-c^{(0)},
\label{A}
\ee
where $\Lambda_a$ is the thermal de Broglie wavelength. This expression is derived using
the regular thermodynamic expression for Helmholtz free energy $A=\sum_aN_a\mu_a-PV$
together with relation
\be
\beta N_a\mu_a=\int\rho_a(1)\left[\ln{\frac{\rho^a_{\left\{0\right\}}(1)}{\Lambda_a}}-
c_{\left\{0\right\}}^a(1)\right]\;\rd(1),
\label{Nmu}
\ee
which follows from (\ref{rhoexp}), written for $\rho^a_{\left\{0\right\}}$. Here $N_a$ is
the number of particles of species $a$ in the system.

Helmholtz free energy in excess to its reference system value $A_\mathrm{ref}$ is
obtained by subtracting corresponding expression for $A_\mathrm{ref}$ from (\ref{A}), i.e.
\be
\beta \left( A-A_\mathrm{ref}\right)=\sum_a\int\left[\rho_a(1)\ln\frac{\rho^a_{\left\{0\right\}}(1)}{\rho_a(1)}+
\sum_{\left\{i\right\}\ne 0}^{\left\{m^a\right\}}
\sigma^a_{\left\{m^a-i\right\}}(1)c^a_{\left\{i\right\}}(1)\right]\;\rd(1)-
\left(c^{(0)}-c^{(0)}_\mathrm{ref}\right),
\label{dA}
\ee
where $c^{(0)}_\mathrm{ref}$ is the
corresponding sum of the diagrams for the reference system. Ordering the virial expansion
(\ref{dA}) with respect to the number of associating $f_{KL}$ bonds and neglecting the
terms with more than one associating bond we have
\be
c^{(0)}-c^{(0)}_\mathrm{ref}=\frac{1}{2}\sum_{ab}\int g_\mathrm{ref}^{ab}(12)\langle {\hat \sigma}_a(1)
{\hat f}_{ab}(12){\hat \sigma}_b(2)\rangle_{12}\;\rd(1)\rd(2)
\label{dc}
\ee
and
\be
{\hat c}_a(1)-c^a_{\left\{0\right\}}(1)=\sum_b\int g_\mathrm{ref}^{ab}(12)\langle
{\hat f}_{ab}(12){\hat \sigma}_a(2)\rangle_2\;\rd(2),
\label{c1}
\ee
where $g_\mathrm{ref}^{ab}(12)$ is the reference system distribution function and
\be
{\hat f}_{ab}(12)=\sum_{KL}\epsilon^a_{K_1}(1)f^{ab}_{KL}(12)
\epsilon^b_{L_1}(2).
\label{fKL}
\ee

Due to
the single bond approximation $c^a_{\left\{i\right\}}=0$ for all values of the set
$\left\{i\right\}$, except for $\left\{i\right\}=0$ and
$\left\{i\right\}=i_K$ with $i_K=1$. This property together with (\ref{rhoexp})
yield the following relations:
\be
c^a_{K_1}(1)=\rho^a_{K_1}(1)/\rho^a_{\left\{0\right\}}(1)
\label{cro}
\ee
and
\be
\frac{\rho^a_{\left\{i\right\}}(1)}{\rho^a_{\left\{0\right\}}(1)}=
\prod_K \frac{1}{i_K!}\left[c^a_{K_1}(1)\right]^{i_K}=
\prod_K \frac{1}{i_K!}\left[\frac{\rho^a_{K_1}(1)}{ \rho^a_{\left\{0\right\}}(1)}\right]^{i_K},
\qquad \text{for} \qquad i_K\in
\left\{i\right\}.
\label{product}
\ee
The set of relations (\ref{dc}), (\ref{c1}) and (\ref{cro}) defined all the quantities
needed to calculate the Helmholtz free energy of the system (\ref{dA}), provided that the
properties of the reference system are known.

Finally it is worth noting, that the ETPT-CF theory developed here
reduces to the TPT1 proposed
by Wertheim~\cite{w34}, if for all sites single-bonding condition $m^a_K=1$ is assumed.
In the other limiting case of only one site per particle the ETPT-CF will coincide with
the TPT-CF developed earlier~\cite{kalstell,dipole,shtock}.

\subsection{Extended TPT-CF for two sites with double-bonding condition}

The theory presented in the previous section is quite general and can be applied to a number
of different situations. However, in the present study we are interested in the version of the
theory for the model with two sites both of which can be bonded twice. More
specifically, we are interested in the extension and application of the
theory to the study of the phase behavior of polydisperse dipolar hard-sphere mixture.

We assume that each of the particles in the system has two doubly bondable attractive sites,
$A$ and $B$, i.e. we have: $M_a=2$ and $m_A^a=m_B^a$. We also assume that attractive interaction
is acting only between the sites of the same sort. Using these suggestions, relations
(\ref{rhoexp}) and (\ref{sigma}), and taking into
account that the system is uniform, the density parameters $\sigma^a_{A_2B_2}=\rho_a$
$\sigma^a_{A_1B_2}\equiv {\tilde \sigma}^a_{A_1}$
and $\sigma^a_{A_2B_1}\equiv{\tilde \sigma}^a_{B_1}a$ can be expressed in terms of $c^a_{K_1}$
\be
\rho_a=\frac{1}{4}\sigma^a_{A_0B_0}\left[1+\left(\kappa^a_{A_1}\right)^2\right]
\left[1+\left(\kappa^a_{B_1}\right)^2\right],
\label{sigmac1}
\ee
\be
{\tilde \sigma}^a_{A_1}=\frac{1}{ 2}\sigma^a_{A_0B_0}\kappa^a_{A_1}
\left[1+\left(\kappa^a_{B_1}\right)^2\right],
\label{sigmac2A}
\ee
\be
{\tilde \sigma}^a_{B_1}=\frac{1}{ 2}\sigma^a_{A_0B_0}\kappa^a_{B_1}
\left[1+\left(\kappa^a_{A_1}\right)^2\right],
\label{sigmac2B}
\ee
where $K$ takes the values $A$ and $B$ and $\kappa^a_{K_1}=1+c^a_{K_1}$.
These two equations give
\be
\sigma^a_{A_0B_0}=4\rho_a\left\{\left[1+\left(\kappa^a_{A_1}\right)^2\right]
\left[1+\left(\kappa^a_{B_1}\right)^2\right]\right\}^{-1},
\label{sigma0}
\ee
and
\be
{\tilde \sigma}^a_{K_1}=\frac{2\rho_a\kappa^a_{K_1}}{ 1+\left(\kappa^a_{K_1}\right)^2}\,.
\label{sigma1}
\ee
In turn, using (\ref{c1}), for $\kappa^a_{K_1}$ we have
\be
\kappa^a_{K_1}=1+\sum_bI_{KK}^{ab}{\tilde\sigma}^a_{K_1},
\label{kappaI}
\ee
where
\be
I_{KK}^{ab}=\int g_\mathrm{ref}^{ab}(12)f^{ab}_{KK}(12)\;\rd(2).
\label{IKK}
\ee
Combining (\ref{dc}), (\ref{sigma1}) and (\ref{kappaI})
the expression for the Helmholtz
free energy (\ref{dA}) can be written in terms
of $\kappa^a_{K_1}$ parameters
\be
\beta\frac{A-A_\mathrm{ref}}{ V}=\sum_a\rho_a\left[\ln{\frac{\rho^a_{A_0B_0}}{ \rho_a}}
-\kappa^a_{A_1}\frac{1-\kappa^a_{A_1}}{ 1+\left(\kappa^a_{A_1}\right)^2}
-\kappa^a_{B_1}\frac{1-\kappa^a_{B_1}}{ 1+\left(\kappa^a_{B_1}\right)^2}\right],
\label{DAfin}
\ee
which satisfy the following set of equations:
\be
\kappa^a_{K_1}=1+\sum_b\frac{2\rho_b\kappa^b_{K_1}
}{ 1+\left(\kappa^b_{K_1}\right)^2}I^{ab}_{KK}\,.
\label{rhoeq}
\ee

Chemical potential $\Delta \mu_a$ and pressure $\Delta P$ in excess of their reference
system values can be obtained using standard thermodynamical relations:
\be
\mu_a-\mu^{a}_\mathrm{ref}=\frac{\partial\left[\left(A-A_\mathrm{ref}\right)/V\right]}
{\partial\rho_a}\,,\qquad
P-P_\mathrm{ref}= \sum_a\rho_a\left(\mu_a-\mu^{a}_\mathrm{ref}\right)-{ \frac{A-A_\mathrm{ref}}{V}}\,.
\label{standard}
\ee

Finally, the average size of the clusters, which appear in the system, can be characterized
by the average length of the chain $L_K$ formed by either $A$-bonded ($K=A$) or $B$-bonded
($K=B$) particles. Following~\cite{duda1,duda2} we defined this quantity by the following expression
\be
L_K={\frac{\sum_a\left(\alpha_{K,\mathrm{end}}^a+\alpha_{K,\mathrm{mid}}^a+\alpha_0^a\right)}{
\sum_a\left(\alpha_{K,\mathrm{end}}^a/2+\alpha_0^a\right)}}\,,
\label{L}
\ee
where $\alpha_{K,\mathrm{end}}^a$ is the fraction of singly $K$-bonded particles
(fraction of the chain ends), $\alpha_{K,\mathrm{mid}}^a$ is the fraction of doubly $K$-bonded
particles (fraction of the chain middles) and $\alpha_0^a$ is the fraction of nonbonded
particles. For these fractions we have
\be
\rho\alpha_{A,\mathrm{end}}^a={\tilde \sigma}_{A_1}^a-\sigma_{A_0B_2}^a\, ,\qquad
\rho\alpha_{A,\mathrm{mid}}^a=\rho-{\tilde \sigma}_{A_1}^a\, ,\qquad
\rho\alpha_0^a=\sigma_{A_0B_0}^a\,.
\label{alpha}
\ee
Substituting these expressions into expression for $L_A$ (\ref{L}) and using
(\ref{sigma0}), (\ref{sigma1}) and expression for $\sigma_{A_0B_2}^a$,
\be
\sigma_{A_0B_2}^a={\frac{1}{2}}\rho_a\sigma_{A_0B_0}\left[\left(\kappa_{B_1}^a\right)^2
+1\right],
\label{B2}
\ee
we get the final expression for $L_A$ in terms of $\kappa_{K_1}^a$:
\be
L_K=\sum_a \rho_a{\frac{4-\left[1+\left(\kappa_{K_1}^a\right)^2\right]
\left[1-\left(\kappa_{{\tilde K}_1}^a\right)^2\right]}{
\left[1+\left(\kappa_{A_1}^a\right)^2\right]
\left[1+\left(\kappa_{B_1}^a\right)^2\right]}}
\left\{\sum_a \rho_a{\frac{4-\left(1-\kappa_{K_1}^a\right)\left[1+\left(\kappa_{{\tilde K}_1}^a\right)^2\right]}{
 \left[1+\left(\kappa_{A_1}^a\right)^2\right]
\left[1+\left(\kappa_{B_1}^a\right)^2\right]}}\right\}^{-1},
\label{LA}
\ee
where if $K=A$ then ${\tilde K}=B$ and if $K=B$ then ${\tilde K}=A$.

\section{Liquid-gas phase behavior of polydisperse dipolar hard-sphere fluid}

\subsection{The model}

We consider a polydisperse dipolar hard-sphere fluid
mixture with a number density $\rho$ and a polydispersity in both
the hard-sphere diameter $\sigma$ and the dipolar moment $d_\mu$.
We assume, that
the dipole moment is proportional to the particle volume, i.e., $d_\mu\sim\sigma^3$.
Thus, the type of the particle is completely defined by its hard-sphere size
and hereafter we will be using $\sigma$ instead of the indices $a,b,\ldots$ to denote
the particle species. We also assume that hard-sphere size of the particles is
distributed according to a normalized distribution function $F(\sigma)\geqslant 0$,
\be
\int_0^\infty F\left(\sigma\right)\;\rd\sigma=1.
\label{F}
\ee
Interaction between particles of species $\sigma_1$ and $\sigma_2$
in our system is described by the following pair potential:
\be
U(r,\sigma_1\sigma_2)=U_\mathrm{hs}(r,\sigma_1\sigma_2)+U_\mathrm{dd}(12,\sigma_1\sigma_2),
\label{Upoly}
\ee
where $U_\mathrm{hs}(r,\sigma_1\sigma_2)$ is the hard-sphere potential and
$U_\mathrm{dd}(r,\sigma_1\sigma_2)$ is the dipole-dipole potential, given by
\be
U_\mathrm{dd}(12,\sigma_1\sigma_2)=-{\frac{d_\mu(\sigma_1)d_\mu(\sigma_2)}{ r^3}}
\left[2\cos{\varphi_1}\cos{\varphi_2}-\sin{\varphi_1}\sin{\varphi_2}
\cos{\left(\phi_1-\phi_2\right)}\right].
\label{Udd}
\ee
Here $\varphi_1$ and $\varphi_2$ denote the angles between the dipole vectors and the vector that joins the centers of the two particles, and $\phi_1$ and $\phi_2$ are the azimuthal
angles about this vector. To proceed we have to split the total potential (\ref{Upoly})
into the reference and the associative pieces. We assume that the reference part of the
potential is represented by the hard-sphere part $U_\mathrm{hs}(r,\sigma_1\sigma_2)$ and the
associative part by the dipole-dipole potential $U_\mathrm{dd}(r,\sigma_1\sigma_2)$. At the contact
distance $\sigma_{12}=(\sigma_1+\sigma_2)/2$, the latter potential has two equal potential
minima of the depth $-2d_{\mu}(\sigma_1)d_{\mu}(\sigma_2)/\sigma_{12}^3$ at ``nose-to-tail''
configuration ($\varphi_1=\varphi_2=0,\;\;\varphi_1=\varphi_2=\pi$). These minima are responsible
for the formation of chains of particles in the system. In addition, there are
twice less deep minima $\left(-d_{\mu}(\sigma_1)d_{\mu}(\sigma_2)/\sigma_{12}^3\right)$ at antiparallel
configuration with $\varphi_1=\varphi_2=\pi/2,\;\;\phi_1-\phi_2=\pi$. The latter minima cause
the formation of the network connecting the chains. According to the earlier theoretical and
computer simulation studies~\cite{tlusty,campPRL} competition between the chain formation and
network formation defines the existence of the liquid-gas phase transition in the dipolar
hard-sphere fluid. To account for this effect we propose the following splitting of the
total associative potential $U_\mathrm{ass}(12,\sigma_1\sigma_2)=U_\mathrm{dd}(12,\sigma_1\sigma_2)$:
\begin{eqnarray}
U_{BB}(12,\sigma_1\sigma_2)&=&\Theta\left(\varphi_1\right)
\Theta\left(\varphi_2\right)U_\mathrm{dd}(12,\sigma_1\sigma_2),
\label{BB}\\
U_{AA}(12,\sigma_1\sigma_2)&=&\left[1-\Theta\left(\varphi_1\right)
\Theta\left(\varphi_2\right)\right]U_\mathrm{dd}(12,\sigma_1\sigma_2),
\label{AA}
\end{eqnarray}
where
$\Theta(\varphi)=H\left(\pi/2+\varphi_0-\varphi\right)H\left(\pi/2-\varphi_0+\varphi\right)$
and $H(x)$ is the Heaviside step function. Here $\varphi_0$ plays a role of the potential
splitting parameter. For $\varphi_0=\pi/2$ we have that
$U_{BB}(12,\sigma_1\sigma_2)=U_\mathrm{dd}(12,\sigma_1\sigma_2)$ and $U_{AA}(12,\sigma_1\sigma_2)=0$.
On the other hand $\varphi_0=0$ gives: $U_{B}(12,\sigma_1\sigma_2)=0$ and
$U_{AA}(12,\sigma_1\sigma_2)=U_\mathrm{dd}(12,\sigma_1\sigma_2)$. In both limiting cases the theory
developed will treat the system as a polydisperse mixture of the hard-sphere chains. For the
intermediate values of $\varphi_0$, the energy minima at ``nose-to-tail'' configuration are
included into $U_{AA}(12,\sigma_1\sigma_2)$ and network forming minima appear in
$U_{BB}(12,\sigma_1\sigma_2)$.
We have chosen here $\varphi_0=\pi/9$. With this value of $\varphi_0$,
our results for monodisperse version of the model are in good agreement with the results of
the previous studies~\cite{campPRL,dipole}.

\subsection{Thermodynamic properties}

For a general multicomponent dipolar hard-sphere mixture, thermodynamic properties can be
obtained using the solution of a set of nonlinear equations (\ref{rhoeq}) and an expression for the Helmholtz free energy (\ref{DAfin}).
However, even for the multicomponent case, solution of this
equation rapidly becomes involved as the number of components increases. As we proceed to
the polydisperse case, solution of the polydisperse version of equation (\ref{rhoeq})
becomes intractable, since now we have to deal with the following integral equation:
\be
\kappa_{K}(\sigma_1)=1+2\rho\int_0^\infty F(\sigma_2){\frac{\kappa_{K}(\sigma_2)
I_{KK}(\sigma_1\sigma_2)}{ 1+\kappa^2_{K}(\sigma_2)}}\;\rd\sigma_2\,,
\label{rhoeqpoly}
\ee
where we have dropped the lower index 1, i.e. $\kappa_{K_1}(\sigma)\equiv \kappa_K(\sigma)$.
In order to solve this equation we propose here to interpolate the key quantity
of the theory, the volume integral $I_{KK}(\sigma_1\sigma_2)$, using a sum of $N_{Y}$ Yukawa
terms. Since the reference system pair distribution function $g_\mathrm{ref}(r,\sigma_1\sigma_2)$
is independent of mutual orientation of the particles for the integral (\ref{IKK}) we have
\be
I_{KK}(\sigma_1\sigma_2)=4\pi\int_0^\infty r^2g_\mathrm{ref}(r,\sigma_1\sigma_2)
{\bar f}_{KK}(r,\sigma_1\sigma_2)\;\rd r,
\label{IKK1}
\ee
where ${\bar f}_{KK}(r,\sigma_1\sigma_2)$ is an orientation averaged Mayer function for
associative potential $U_{KK}(12,\sigma\sigma)$. We assume, that
${\bar f}_{KK}(r,\sigma_1\sigma_2)$ can be represented in the following form:
\be
{\bar f}_{KK}(r,\sigma_1\sigma_2)={\frac{1}{ 4\pi r}}\sum_n^{N_Y}
A^{(n)}_K(\sigma_1)A^{(n)}_K(\sigma_2)
\re^{-z_K^{(n)}\left(r-\sigma_{12}\right)}.
\label{barf}
\ee
Parameters $A^{(n)}_K(\sigma)$ and $z^{(n)}_K$ are obtained using the
interpolation scheme, which is presented and discussed in the appendix~A.
Using (\ref{IKK1}) and (\ref{barf}), we have
\be
I_{KK}(\sigma_1\sigma_2)=\sum_n^{N_Y}A^{(n)}_K(\sigma_1)A^{(n)}_K(\sigma_2)
G_{K}^{(n)}(\sigma_1\sigma_2),
\label{IKK2}
\ee
where
$G_{K}^{(n)}(\sigma_1\sigma_2)=\int_0^\infty r\re^{-z_K^{(n)}r}g_\mathrm{ref}(r,\sigma_1\sigma_2)\rd r$
is the Laplace transform of the radial
distribution function $g_\mathrm{ref}(r,\sigma_1\sigma_2)$.
We will use here Percus-Yevick approximation for the hard-sphere radial distribution
function, since the analytical expression for its Laplace transform is known
\cite{blum}
\begin{eqnarray}
\re^{-z^{(n)}_K\sigma_{12}}G^{(n)}_K(\sigma_1\sigma_2)&=&\frac{ \Delta
}{(z^{(n)}_K)^2D_K^{(n)}}\Bigg[z_K^{(n)}\left(\sigma_{12}
+\sigma_{1}\sigma_{2}\frac{\pi}{4\Delta}m_2\right)+1
+\frac{\pi}{2\Delta}m_3\nonumber\\
&&+\ \frac{\pi{}z_K^{(n)}}{2\Delta}\left(m_{K,2}^{(n)}-2\sigma_{12}m_{K,1}^{(n)}
+\sigma_1\sigma_2m_{K,0}^{(n)}\right)\Bigg],
\label{gs}
\end{eqnarray}
where
\begin{eqnarray}
D_K^{(n)}&=&\Delta^2-{\frac{2\pi}{ z_K^{(n)}}}\left(1+{\frac{1}{ 2}}\pi m_3\right)
\left(m_{K,0}^{(n)}+{\frac{1}{2}}m_2\right)\nonumber\\
&&-\ 2\pi\left\{\Delta m_{K,1}^{(n)}+{\frac{1}{ 4}}\pi
\left[m_{K,2}^{(n)}\left(m_2+2m_{K,0}^{(n)}\right)-\left(m_{K,1}^{(n)}\right)^2\right]\right\},
\label{D0s}
\end{eqnarray}
\be
\Delta=1-\pi m_3/6.
\label{delta}
\ee
Here $m_l$ are the moments and $m^{(n)}_{K,l}$ are the generalized moments of the distribution
function $F(\sigma)$.  Expression for these moments can be symbolically presented as follows:
\be
m=\int_0^\infty m(\sigma)F(\sigma)\;\rd\sigma.
\label{m}
\ee
Hereafter all the quantities denoted as $m$ with certain set of indices will represent the
generalized moments defined by~(\ref{m}). Corresponding expressions for $m(\sigma)$ are
collected in the appendix B.
Inserting~(\ref{IKK2}) into equation (\ref{rhoeqpoly}) and using (\ref{gs}), we find
\be
\kappa_K(\sigma)=1+\rho\sum_n^{N_Y}\sum_{j=1}^2C_{K,j}^{(n)}\Omega_{K,j}^{(n)}(\sigma),
\label{solution}
\ee
where $C_{K,j}^{(n)}$ satisfies the following set of equations:
\be
C^{(n)}_{K,j}=2\int_0^\infty\sigma^{j-1}F(\sigma)A_K^{(n)}(\sigma)
{\frac{1+\rho\sum_l\sum_{i=1}^2\Omega^{(l)}_{K,i}(\sigma)C^{(l)}_{K,i}}{
1+\left[1+\rho\sum_l\sum_{i=1}^2\Omega^{(l)}_{K,i}(\sigma)C^{(l)}_{K,i}\right]^2}}\;\rd\sigma.
\label{Ceq}
\ee
Here
\be
\Omega_{K,1}^{(n)}(\sigma)=A_K^{(n)}(\sigma)\left(\sigma P_{K,1}^{(n)}+P_{K,3}^{(n)}\right),
\qquad
\Omega_{K,2}^{(n)}(\sigma)=A_K^{(n)}(\sigma)\left(\sigma P_{K,2}^{(n)}+P_{K,1}^{(n)}\right),
\label{QQ}
\ee
\begin{eqnarray}
P_{K,1}^{(n)}&=&{\frac{\Delta}{ 2z_K^{(n)}D^{(n)}_K}}\left(1-{\frac{\pi}{ \Delta}}m_{K,1}^{(n)}\right),
\label{P1}\\
P_{K,2}^{(n)}&=&{\frac{\pi}{ 2z_K^{(n)}D^{(n)}_K}}\left({\frac{1}{2}}m_2+m_{K,0}^{(n)}\right),
\label{K2}\\
P_{K,3}^{(n)}&=&{\frac{\Delta}{ \left(z_K^{(n)}\right)^2D^{(n)}_K}}
\left[1+{\frac{\pi}{ 2\Delta}}\left(m_3+z^{(n)}_Km^{(n)}_{K,2}\right)\right].
\label{P3}
\end{eqnarray}
Thus, solution of the integral equation (\ref{rhoeqpoly}) for the unknown function
$\kappa_K(\sigma)$ now is reduced to the solution of a set of equations for $4N_Y$ unknown
constants $C_{K,j}^{(n)}$. This solution can be used to calculate
$\kappa_K^{(n)}(\sigma)$, which in turn can be utilized to calculate thermodynamical
properties of the system via Helmholtz free energy (\ref{DAfin}). Generalizing
the expression for Helmholtz free energy (\ref{DAfin})
for a polydisperse system, we have
\be
\beta{\frac{A-A_\mathrm{ref}}{ V}}=\rho\int_0^\infty F(\sigma)\left\{-\ln{{\frac{1}{ 4}}\prod_{K=A}^B
\left[\kappa_A^2(\sigma)+1\right]}
-\sum_{K=A}^B\kappa_K(\sigma){\frac{1-\kappa_K(\sigma)}{ 1+\kappa_K^2(\sigma)}}\right\}
\;\rd\sigma.
\label{DAfinpoly}
\ee
Now we can use the standard relation between Helmholtz free energy
and chemical potential
(\ref{standard}), generalized to polydisperse case
\be
\beta\left[\mu(\sigma)-\mu_\mathrm{ref}(\sigma)\right]=
{\frac{\beta}{ \rho}}\;{\frac{\delta\left\{A-A_\mathrm{ref}/V\right\}}{\delta\left\{F(\sigma)\right\}}}\,,
\label{standardfun}
\ee
where $\delta/\delta\left\{F(\sigma)\right\}$ denote functional differentiation with respect to
the distribution $F(\sigma)$. We find
\begin{eqnarray}
\beta\left[\mu(\sigma)-\mu_\mathrm{ref}(\sigma)\right]&=&{\frac{m_\mu}{ \rho}}
-\ln{{\frac{1}{ 4}}\prod_{K=A}^B
\left[\kappa_A^2(\sigma)+1\right]}-
\sum_{K=A}^B\kappa_K(\sigma){\frac{1-\kappa_K(\sigma)}{ 1+\kappa_K^2(\sigma)}}\nonumber\\
&&+\ \sum_n^{N_Y}\sum_{K=A}^B\left(\sum_{j-1}^3\mu_{K,j}^{(n)}{\frac{\delta P_{K,j}^{(n)}}{F(\sigma)}}
+\sum_{j=1}^2\nu_{K,j}^{(n)}{\frac{\delta C_{K,j}^{(n)}}{ F(\sigma)}}\right).
\label{mufun}
\end{eqnarray}
Here
\begin{eqnarray}
\mu_{K,1}^{(n)}&=&{\frac{1}{2}}\left(m_{K,0}^{(n,0)}C_{K,2}^{(n)}+m_{K,1}^{(n,0)}C_{K,1}^{(n)}\right)\,,
\label{muK1}\\
\mu_{K,2}^{(n)}&=&{\frac{1}{2}}\left(m_{K,1}^{(n,0)}-\rho C_{K,2}^{(n)}\right)C_{K,2}^{(n)}\,,
\qquad
\mu_{K,3}^{(n)} \quad = \quad {\frac{1}{2}}\left(m_{K,0}^{(n,0)}-\rho C_{K,1}^{(n)}\right)C_{K,1}^{(n)}\,,
\label{muK23}\\
\nu_{K,1}^{(n)}&=&{\frac{1}{ 2}}\left[\left(m_{K,1}^{(n,0)}-\rho C_{K,2}^{(n)}\right)P_{K,1}^{(n)}+
\left(m_{K,0}^{(n,0)}-\rho C_{K,1}^{(n)}\right)P_{K,3}^{(n)}\right]\,,
\label{nu1}\\
\nu_{K,2}^{(n)}&=&{\frac{1}{ 2}}\left[\left(m_{K,0}^{(n,0)}-\rho C_{K,1}^{(n)}\right)P_{K,1}^{(n)}+
\left(m_{K,1}^{(n,0)}-\rho C_{K,2}^{(n)}\right)P_{K,2}^{(n)}\right]\,,
\label{nu2}
\end{eqnarray}
functional derivatives $\delta P_{K,j}^{(n)}/ \delta F(\sigma)$ and
$\delta D_K^{(n)}/ \delta F(\sigma)$ are presented in the appendix~B
and functional derivatives $\delta C_{K,j}^{(n)}/ \delta F(\sigma)$
can be obtained from the solution of the set of linear equations,
which follows from (\ref{Ceq}) upon its functional
differentiation with respect to $F(\sigma)$, i.e.
\be
\sum_{j=1}^2{\bf M}_{i,j}^{(K)}{\frac{\delta {\bf C}_j^{(K)}}{ \delta F(\sigma)}}=
{\bf R}_i^{(K)}(\sigma),
\qquad i=1,2,\qquad K=A,B,
\label{dCeq}
\ee
where $\left[\delta {\bf C}_j^{(K)}/ \delta F(\sigma)\right]_n
\equiv\delta C_{K,j}^{(n)}/ \delta F(\sigma)$
and the elements of the matrices ${\bf M}_{i,j}^{(K)}$ and
${\bf R}_i^{(K)}(\sigma)$ are collected in the appendix~B.

The pressure expression follows from (\ref{standard}), generalized to
the polydisperse case
\be
P-P_\mathrm{ref}= \rho\int_0^\infty F(\sigma)\left[\mu(\sigma)-\mu_\mathrm{ref}(\sigma)
\right]\rd\sigma-{\frac{A-A_\mathrm{ref}}{ V}}\,.
\label{Pfun}
\ee
Using this expression together with the expression for the chemical potential
(\ref{mufun}), we find
\be
\beta\left(P-P_\mathrm{ref}\right)=m_\mu+\rho\sum_n^{N_Y}\sum_{K+A}^B\left(\sum_{j=1}^3
\mu_{K,j}^{(n)}J_{K,j}^{(n)}+\sum_{j=1}^2\nu_{K,j}^{(n)}S_{k,j}^{(n)}\right),
\label{Pfun1}
\ee
where
\be
J_{K,j}^{(n)}=\int_0^\infty F(\sigma){\delta P_{K,j}^{(n)}\over \delta F(\sigma)}\rd\sigma,
\qquad
S_{K,j}^{(n)}=\int_0^\infty F(\sigma){\delta C_{K,j}^{(n)}\over \delta F(\sigma)}\rd\sigma.
\label{JS}
\ee
Expression for the integral $J_{K,j}^{(n)}$ is presented in the appendix~B
and integral $S_{K,j}^{(n)}$ can be obtained from the solution of the set of
linear equations
\be
\sum_{j=1}^2{\bf M}_{i,j}^{(K)}{\bf S}_j^{(K)}=
{\bf E}_i^{(K)}\,,
\qquad i=1,2, \qquad K=A,B,
\label{intCeq}
\ee
which follows from the set of equations (\ref{dCeq}). Here
$\left[{\bf S}_j^{(K)}\right]_n\equiv S_{K,j}^{(n)}$
and the elements of the matrix ${\bf E}^{(K)}_i$ are presented in the appendix~B.

Expressions for the chemical potential (\ref{mufun}) and pressure (\ref{Pfun1})
are the final expressions to be used in the phase equilibrium calculations. The
properties of the reference system (chemical potential $\mu_\mathrm{ref}(\sigma)$ and
pressure $P_\mathrm{ref}$) are described
here using polydisperse versions of the Mansoori et al.~\cite{mansoori}
expressions
\begin{eqnarray}
\beta\mu^{(\mathrm{ex})}_\mathrm{ref}(\sigma)&=&\left[\left(\sigma{m_2\over m_3}\right)^2
\left(3-2\sigma{m_2\over m_3}\right)-1\right]\ln{\Delta}
+m_2\left(1+{m_2\sigma\over m_3\Delta}\right)\nonumber\\
&& \ +{\pi\sigma\over 2\Delta}
\left\{{1\over 3}\sigma^2\left[\rho-{m_2^3\over m_3^2}{1+\Delta\over \Delta}
+{\pi\over \Delta}m_2\left({\frac{1}{ 2}}m_1+{1\over 3}{m_2^2\over m_3\Delta}
\right)\right]+\sigma m_1\right\},
\label{muhs}\\
\beta P_\mathrm{ref}&=&{1\over \Delta}\left[\rho+{\pi\over 2\Delta}m_1m_2+{\pi^2\over 12\Delta^2}m_2^3
-\left({\pi\over 6}\right)^3{1\over \Delta^2}m_3m_2^3\right],
\label{Phs}
\end{eqnarray}
where $\mu_\mathrm{ref}^{(\mathrm{ex})}(\sigma)$ is the reference system chemical potential in excess
to its ideal gas value.

\subsection{Phase equilibrium conditions}

One can easily see that thermodynamical properties of the model at hand obtained above
are defined by the set of the finite number of the distribution function moments, i.e.,
four regular moments ($m_l$, $l=0,1,2,3$) and $1+10N_Y+3N_Y\left(N_Y+1\right)$ generalized moments
($m_\mu,\;m_{K,i}^{(n)}\;m_{K,j}^{(n0)}\;m_{K,l}^{(nm)};\;i=0,1,2;\;j=0,1;\;l=0,1,2;\;K=A,B$).
Note, that $m_{K,l}^{(nm)}=m_{K,l}^{(mn)}$. Thus, the polydisperse mixture
of dipolar hard spheres treated within ETPT-CF belongs to the class of truncatable
free energy models (see~\cite{sollich} and references therein). This property allows us to
map the phase coexistence relations onto a set of nonlinear equations
for the unknown moments
of the daughter distribution functions~\cite{xu}.

We assume that at a certain density $\rho_0$ and composition $F_0(\sigma)$
the system separates into two phases with densities $\rho_1$ and $\rho_2$,
and compositions $F_1(\sigma)$ and $F_2(\sigma)$. Hereafter the lower
index $0$ refers to the parent phase and the lower indices $1$ and $2$
refer to the daughter phases. At equilibrium these quantities take the values, which
follows from the phase equilibrium conditions, i.e.: (i) conservation of the
total volume of the system, (ii) conservation of the total number of the particles of
each species, (iii) equality of the chemical potentials of particles of the same
species in the coexisting phases, (iv) equality of the pressure in the coexisting
phases. These conditions finally lead to the following set of relations~\cite{xu,kal1}:
\be
F_\alpha(\sigma)=F_0\left(\sigma\right)
Q_\alpha\left(\sigma;\rho_0,
\rho_1,\rho_2;\left[F_\alpha\right]\right),
\label{FQ}
\ee
\be
P_1\left(\rho_1;\left[F_1\right]\right)=
P_2\left(\rho_2;\left[F_2\right]\right),
\label{P1P2}
\ee
\be
\int_0^\infty F_\alpha(\sigma)\;d\sigma=1,
\qquad \text{for} \quad \alpha=1 \quad \text{or} \quad \alpha=2,
\label{norm12}
\ee
where
\begin{eqnarray}
\rho_\alpha Q_\alpha\left(\sigma;\rho_0,\rho_1,\rho_2;
\left[F_\alpha\right]\right)&=&
{\rho_0\left(\rho_2-\rho_1\right)\left[1-\delta_{1\alpha}+
\delta_{1\alpha}
\exp \left(\beta\Delta\mu^{(\mathrm{ex})}\right)
\right]\over
\rho_0-\rho_1-\left(\rho_0-\rho_2\right)
\exp \left(\beta\Delta\mu^{(\mathrm{ex})}\right)}\,,
\label{Q}\\
\Delta\mu^{(\mathrm{ex})}
&=&\mu^{(\mathrm{ex})}_2\left(\sigma,\rho_2;\left[F_2\right]\right)
-\mu^{(\mathrm{ex})}_1\left(\sigma,\rho_1;\left[F_1\right]\right),
\label{Dmu}
\end{eqnarray}
$\mu_\alpha^{(\mathrm{ex})}$ is the excess (over the ideal gas) chemical potential
of the particle of species $\sigma$ in the phase $\alpha$, and $\left[\ldots\right]$
denote functional dependence. The relation between $F_0(\sigma)$ and daughter
phase distribution function $F_\alpha(\sigma)$, i.e., equation~(\ref{FQ}), follows
from the phase equilibrium conditions (i)--(iii).

Relations (\ref{FQ})--(\ref{norm12}) represent a closed set of equations to be solved
for the unknowns $\rho_\alpha$ and $F_\alpha(\sigma)$; this set has to
be solved for every value of the species variable $\sigma$. However,
since thermodynamical properties of the model at hand are defined by the finite number
of the moments we can map this set of equations onto a closed set of
$10+28N_Y+6N_Y\left(N_Y+1\right)$
algebraic equations for $\rho_\alpha$, $C_{K,1}^{(n)(\alpha)},\;C_{k,2}^{(n)(\alpha)}$
and moments
$m^{(\alpha)}_k,\;m^{(\alpha)}_\mu,\;m_{K,i}^{(n)(\alpha)},\;m_{K,j}^{(n0)(\alpha)},\;
m_{K,l}^{(nm)(\alpha)}$ in the two coexisting phases ($\alpha=1,2$).
We have
\begin{eqnarray}
m_k^{(\alpha)}&=&\rho_\alpha\int_0^\infty m_k(\sigma)F_0\left(\sigma\right)
Q_\alpha\left(\sigma,\rho_0;\left\{X_1\right\},\left\{X_2\right\}\right)\;\rd\sigma,
\qquad k=0,1,2,3,
\label{m1}\\
m_\mu^{(\alpha)}&=&\rho_\alpha\int_0^\infty m^{(\alpha)}_\mu(\sigma)F_0\left(\sigma\right)
Q_\alpha\left(\sigma,\rho_0;\left\{X_1\right\},\left\{X_2\right\}\right)\;\rd\sigma,
\label{m2}\\
m_{K,i}^{(n)(\alpha)}&=&\rho_\alpha\int_0^\infty m_{K,i}^{(n)(\alpha)}(\sigma)F_0\left(\sigma\right)
Q_\alpha\left(\sigma,\rho_0;\left\{X_1\right\},\left\{X_2\right\}\right)\;\rd\sigma,
\qquad i=0,1,2,
\label{m3}\\
m_{K,j}^{(n0)(\alpha)}&=&\rho_\alpha\int_0^\infty m_{K,j}^{(n0)(\alpha)}(\sigma)F_0\left(\sigma\right)
Q_\alpha\left(\sigma,\rho_0;\left\{X_1\right\},\left\{X_2\right\}\right)\;\rd\sigma,
\qquad j=0,1,
\label{m4}\\
m_{K,l}^{(nm)(\alpha)}&=&\rho_\alpha\int_0^\infty m_{K,l}^{(nm)(\alpha)}(\sigma)F_0\left(\sigma\right)
Q_\alpha\left(\sigma,\rho_0;\left\{X_1\right\},\left\{X_2\right\}\right)\;\rd\sigma,
\qquad l=0,1,2,
\label{m5}
\end{eqnarray}
where $K=A,B$ and
$\left\{X_\alpha\right\}$ represent the unknowns of the problem, i.e.
$$
\left\{X_\alpha\right\}=
\left\{\rho_\alpha,m^{(\alpha)}_k,\;m^{(\alpha)}_\mu,\;m_{K,i}^{(n)(\alpha)},\;m_{K,j}^{(n0)(\alpha)},\;
m_{K,l}^{(nm)(\alpha)}\right\},
\qquad \alpha=1,2.
$$
The remaining $2+8N_Y$ equations follow from the equality of the pressure in coexisting
phases (\ref{P1P2}),
\be
P_1\left(\rho_1;\left\{X_1\right\}\right)=
P_2\left(\rho_2;\left\{X_2\right\}\right),
\label{P1P2X}
\ee
from the set of equations for $C_{K,1}^{(n)(\alpha)}$ and $C_{K,2}^{(n)(\alpha)}$
(\ref{Ceq}) and from the normalizing condition (\ref{norm12}) for either phase
$\alpha=1$ or $\alpha=2$,
\be
\int_0^\infty F_0\left(\sigma\right)
Q_\alpha\left(\sigma,\rho_0;
\left\{X_1\right\},\left\{X_2\right\}\right)\;\rd\sigma=1.
\label{norm}
\ee
Solution of the set of equations (\ref{Ceq}), (\ref{m1})--(\ref{norm}) for a given density
$\rho_0$ and distribution function $F_0(\sigma)$ of the parent phase
gives the densities $\rho_\alpha$ and distribution functions
$F_\alpha(\sigma)$ of the two coexisting daughter phases. The coexisting
densities at different densities of the parent phase $\rho_0$ defined binodals,
which are terminated when the density of one
of the phases is equal to the parent phase
density $\rho_0$. These termination points form cloud and shadow coexisting
curves. These curves intersect at the critical point, which is characterized by the
critical density $\rho_\mathrm{cr}=\rho_1=\rho_2=\rho_0$ and critical temperature $T_\mathrm{cr}$.
The cloud-shadow curves can be obtained as a special solution of the general
coexisting problem, when the properties of one phase are equal to the properties
of the parent phase: assuming that the phase $\alpha=2$ is the cloud phase, i.e.
$\rho_2=\rho_0$, and following the above scheme we will end up with the
same set of equations (\ref{Ceq}), (\ref{m1})--(\ref{norm}), but with $\rho_2$ and
$F_2(\sigma)$ substituted by $\rho_0$ and $F_0(\sigma)$,
respectively.

\section{Results and discussion}

In this section we present our numerical results for a liquid-gas phase diagram of
polydisperse dipolar hard-sphere fluid at different degrees of polydispersity. For
a size distribution function $F(\sigma)$ we have chosen the beta distribution given by
\be
F(\sigma)={\Gamma\left(\nu+\mu-1\right)\over \Gamma\left(\nu\right)
\Gamma\left(\mu\right)}
{\left(1-x\right)^{\mu-1}x^{\nu-1}\over \left(\sigma_u-\sigma_d\right)}\;
H\left(\sigma_u-\sigma\right)H\left(\sigma-\sigma_d\right),
\label{Fbeta}
\ee
where
\be
x={\sigma-\sigma_d\over \sigma_u-\sigma_d}\,,\qquad
\nu={1-{\tilde \sigma}_0\over D_\sigma}-{\tilde \sigma}_0\,, \qquad
\mu=\left({1\over {\tilde \sigma}_0}-1\right)\nu,\qquad
{\tilde \sigma}_0={\sigma_0-\sigma_d\over \sigma_u-\sigma_d}\,,
\label{x1}
\ee
\be
\sigma_0=\langle\sigma\rangle,\qquad
D_\sigma={\langle\sigma^n\rangle\over \sigma_0^2}-1,\qquad
\langle\sigma^n\rangle=\int\;\rd\sigma^n \sigma F(\sigma).
\label{x2}
\ee
Here $\sigma_u$ and $\sigma_d$ define the range of values for $\sigma$.
In our calculations we have chosen
$\sigma_u=1.2947\sigma_0$ and $\sigma_d=0.85\sigma_0$.

\begin{figure}[ht]
\centerline{\includegraphics[clip,width=85mm]{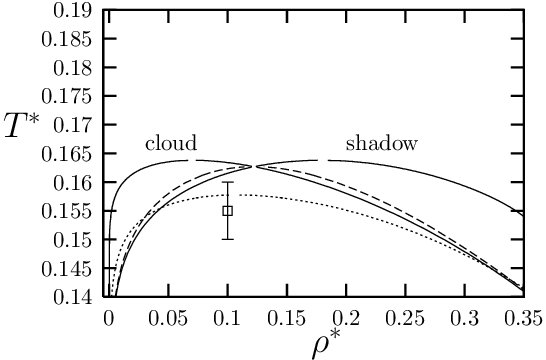}}
\caption{Predictions of the ETPT-CF for the phase diagram
of polydisperse dipolar hard-sphere
mixture including cloud and shadow curves (as labeled),
and critical binodal (dashed line) at $D_\sigma=0.1$
in $\rho^*\;\text{vs}\;T^*$ coordinate frame.
Dotted line and empty square represent ETPT-CF binodal
and MC~\cite{campPRL} critical
point of monodisperse dipolar hard-sphere fluid, respectively.}
\label{f1}
\end{figure}

\begin{figure}[ht]
\centerline{\includegraphics[clip,width=0.49\textwidth]{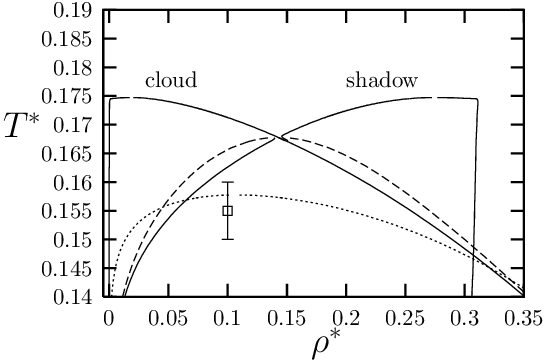}
\hfill
\includegraphics[clip,width=0.49\textwidth]{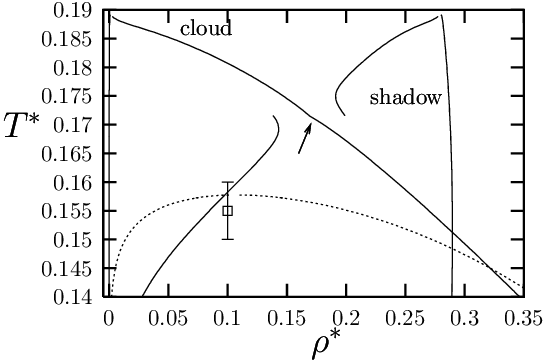}}
\centerline{
\parbox[t]{0.48\textwidth}{\caption{The same as in figure~\ref{f1} at $D_\sigma=0.2$.}
\label{f2}}
\hfill
\parbox[t]{0.48\textwidth}{\caption{The same as in figure~\ref{f1} at $D_\sigma=0.3$.}
\label{f3}}}
\end{figure}

\begin{figure}[ht]
\centerline{\includegraphics[clip,width=85mm]{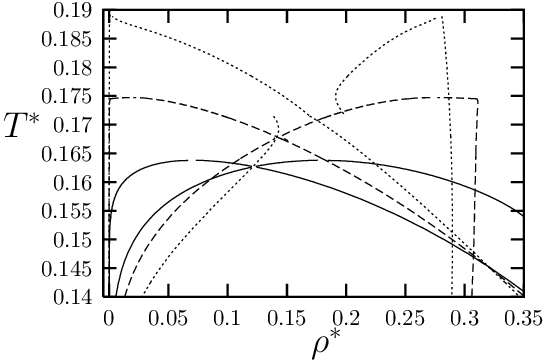}}
\caption{ETPT-CF cloud and shadow curves for polydisperse dipolar
hard-sphere mixture at $D_\sigma=0.1$ (solid lines),
$D_\sigma=0.2$ (dashed lines) and $D_\sigma=0.3$ (dotted lines).}
\label{f4}
\end{figure}

\begin{figure}[t]
\centerline{
\includegraphics[clip,width=0.49\textwidth]{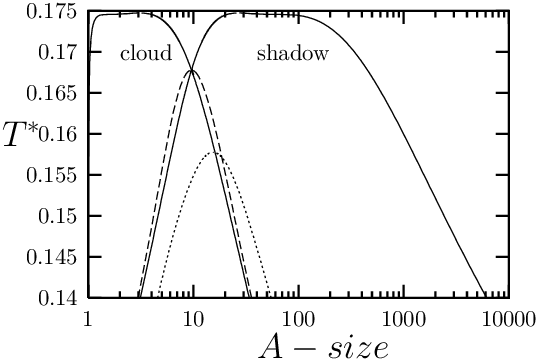}
\hfill
\includegraphics[clip,width=0.49\textwidth]{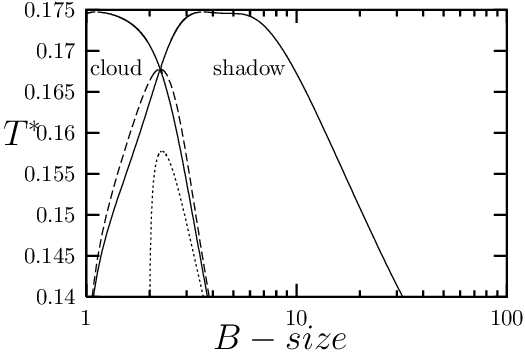}}
\centerline{
\parbox[t]{0.48\textwidth}{\caption{Average $A$-size of the clusters along the cloud and shadow curves
(as labeled), along the critical binodal (dashed line) for polydisperse
dipolar hard-sphere fluid at $D_\sigma=0.2$
and along the binodal for monodisperse version of the model (dotted lines).}
\label{f5}}
\hfill
\parbox[t]{0.48\textwidth}{\caption{Average $B$-size of the clusters along the cloud and shadow curves
(as labeled), along the critical binodal (dashed line) for polydisperse
dipolar hard-sphere fluid at $D_\sigma=0.2$
and along the binodal for monodisperse version of the model (dotted lines).}
\label{f6}}
}
\end{figure}

\begin{figure}[ht]
\centerline{\includegraphics[clip,width=85mm]{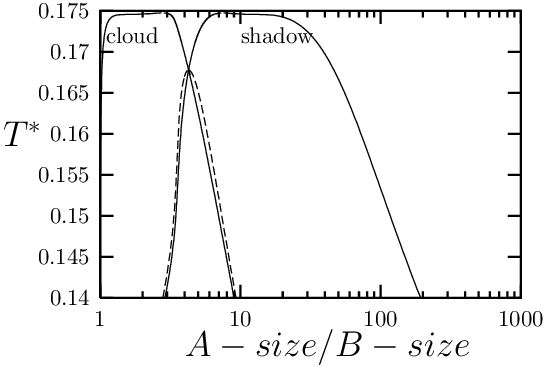}}
\caption{The ratio of the average A and B sizes of the clusters
along the cloud and shadow curves
(as labeled), along the critical binodal (dashed line) for polydisperse
dipolar hard-sphere fluid at $D_\sigma=0.2$.}
\label{f7}
\end{figure}

\begin{figure}[ht]
\centerline{\includegraphics[clip,width=85mm]{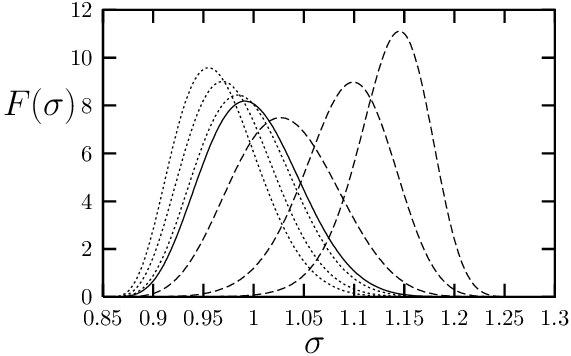}}
\caption{Distribution functions of the gas (dotted lines) and liquid
(dashed lines) phases along the shadow curve for $T^*=0.16,\;0.15,
\;0.14$ and mother phase distribution function (solid line)
at $D_\sigma=0.1$. With the temperature decrease distribution functions
of the liquid phase shifts in the direction of larger $\sigma$ and
distribution functions of the gas phase shifts in the direction of
smaller $\sigma$.}\label{f8}
\end{figure}

\begin{figure}[ht]
\centerline{
\includegraphics[clip,width=0.49\textwidth]{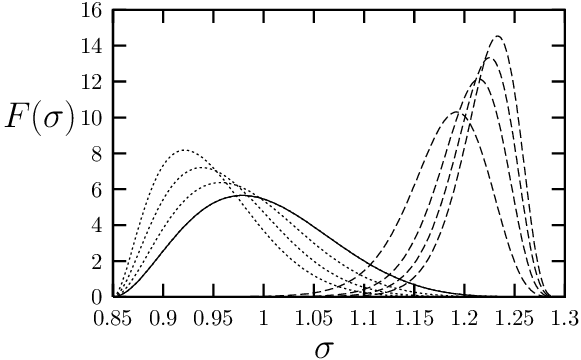}
\hfill
\includegraphics[clip,width=0.49\textwidth]{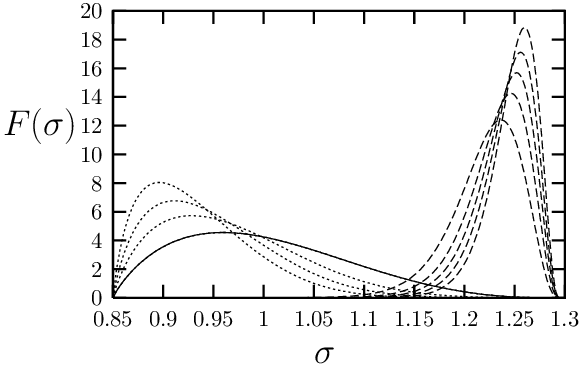}}
\centerline{
\parbox[t]{0.48\textwidth}
{\caption{The same as in figure~\ref{f8} for $T^*=0.17,\;0.16,
\;0.15,\;0.14$ and $D_\sigma=0.2$.}\label{f9}
}
\hfill
\parbox[t]{0.48\textwidth}{\caption{The same as in figure~\ref{f8} for $T^*=0.18,\;0.17,\;0.16,\;0.15,\;0.14$
 and $D_\sigma=0.3$.}\label{f10}
}}
\end{figure}

\begin{figure}[!ht]
\centerline{\includegraphics[clip,width=85mm]{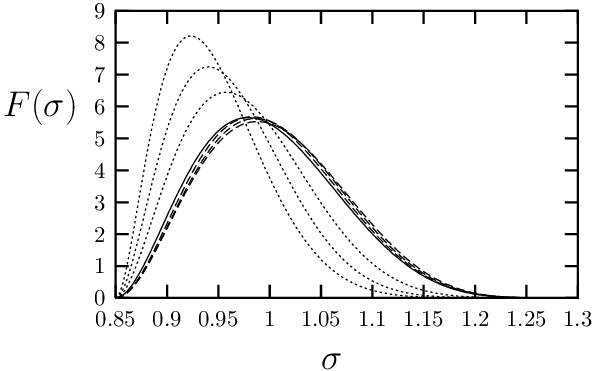}}
\caption{Distribution functions of the gas (dotted lines) and liquid
(dashed lines) phases along the critical binodal for $T^*=0.16,\;0.15,\;0.14$
and mother phase distribution function (solid line)
at $D_\sigma=0.2$. With the temperature decrease, distribution functions
of the liquid phase slightly shifts in the direction of larger $\sigma$
(almost coinciding with the mother phase distribution function) and
distribution functions of the gas phase shifts in the direction of
smaller $\sigma$.}\label{f11}
\end{figure}

\begin{figure}[ht]
\centerline{\includegraphics[clip,width=85mm]{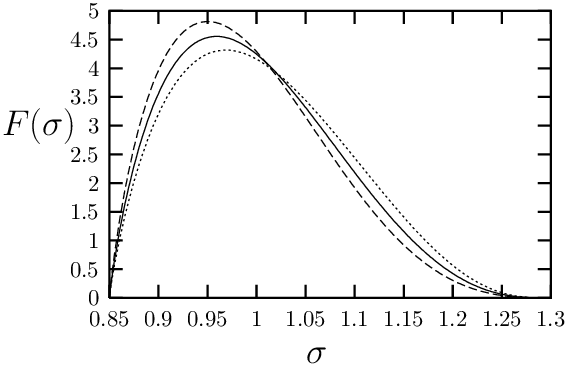}}
\caption{Distribution functions on a shadow curves
at $T^*=0.1715$ for three phases at equilibria for $D_\sigma=0.3$.}\label{f12}
\end{figure}

In figures~\ref{f1}--\ref{f3} the liquid-gas phase diagram for polydisperse dipolar hard-sphere
fluid at different degrees of polydispersity $D_\sigma\!=0.1,\;0.2,\;0.3$ is
presented in $T^*\text{vs}\;\rho^*$ coordinate frame. Here $T^*\!\!=k_\mathrm{B}T/d^2_\mu(\sigma_0)$
and $\rho^*=\rho \sigma_0^3$.
We show the cloud and shadow curves and critical binodal. In addition,
for the reference we include Monte-Carlo predictions for the critical point
and ETPT-CF predictions for the phase diagram of monodisperse version of the model~\cite{campPRL}. One can see that upon increasing the $D_\sigma$, the region
of the phase instability increases with the critical point shifting to higher
temperatures and densities. For larger values of $D_\sigma$ ($D_\sigma=0.2,\;0.3$),
the low density part of the cloud curve and the high density part of the shadow curve
become almost vertical. For $D_\sigma=0.3$ the cloud and shadow curves do not intersect.
At $T^*=0.1715$ the cloud curve has a cusp (denoted by the arrow in figure~\ref{f3})
and shadow curve has a jump discontinuity.
We believe that at this temperature there is a three-phase equilibria, when the mother
phase is in equilibrium with two phases on two branches of the shadow curve, one with
slightly lower density and the other with slightly higher density, respectively.
The cloud and shadow curves for the whole set of values for $D_\sigma$ are collected in
figure~\ref{f4}.
In figures~\ref{f5} and \ref{f6} we present the average $A$-size $L_A$ and $B$-size
$L_B$ of the clusters, respectively, along the
cloud and shadow curves and along the binodals for both monodisperse and polydisperse
(with $D_\sigma=0.2$) versions
of the model. For $L_K$ we have used the expression (\ref{LA}), extended to account for
polydispersity, i.e.
\be
L_K=\int \rd\sigma F(\sigma){4-\left[1+\kappa^2_{K_1}(\sigma)\right]
\left[1-\kappa^2_{{\tilde K}_1}(\sigma)\right]\over
\left[1+\kappa^2_{A_1}(\sigma)\right]
\left[1+\kappa^2_{B_1}(\sigma)\right]}
\left\{\int \rd\sigma F(\sigma){4-\left(1-\kappa_{K_1}(\sigma)\right)\left[1+\kappa^2_{{\tilde K}_1}(\sigma)\right]
\over \left[1+\kappa^2_{A_1}(\sigma)\right]
\left[1+\kappa^2_{B_1}(\sigma)\right]}\right\}^{-1}.
\label{LApoly}
\ee
From these figures one can see that clusters of larger sizes occur in the liquid phase, in
comparison with the gas phase. A decrease of the temperature causes an increase of the cluster
sizes in the liquid phase and a decrease in the gas phase. With an increase of polydispersity,
$L_K$ along the cloud curve does not change much. However, corresponding changes along the
shadow curve are more substantial, here the cluster sizes increase with an increase of polydispersity. In figure~\ref{f7} we show the ratio of $A$- and $B$-sizes $L_A/L_B$. As one would
expect $L_A$ is substantially larger than $L_B$, thus a chain structure of the
formed clusters prevails, with chains mutually connected via $B$-bonds. This difference in $L_A$
and $L_B$ is larger in the liquid phase and increases with polydispersity increase. Similarly,
as before, the temperature decrease causes $L_A/L_B$ decrease in the gas phase and increase
in the liquid phase. Figures~\ref{f8}--\ref{f10} and \ref{f11} show distribution functions of the shadow curve
and on the critical binodal at different temperatures. According to these
figures the larger size particles always fractionate to the liquid phase and smaller particles
tend to move to the gas phase. With an increase of $D_\sigma$ and a decrease of the temperature,
these fractionation effects become more pronounced. Finally in figure~\ref{f12} we show distribution
functions of the two branches of the shadow curve at $T^*=0.1715$ and mother phase distribution
function.

\vspace{0.5cm}

\vspace{0.5cm}

\section{Conclusions}

In this paper we propose an extension of our TPT-CF approach to account for several associating
sites with the possibility of each site to be multiply bonded. The theory is applied
to the study of the liquid-gas phase behavior of a polydisperse dipolar
hard-sphere fluid with polydispersity in both hard-sphere size and dipole moment.
It is assumed that the dipole moment is proportional
to the volume of the particle.
We present a full phase diagram, which includes cloud and shadow curves, binodals
and distribution functions of coexisting
phases and discuss the effects of  polydispersity on their behavior.
According to our analysis,  polydispersity extends the region of the phase instability
shifting the critical point to higher values of temperature and density.
For lower values of temperature, polydispersity causes strong
fractionation effects, with the larger size particles always tending to the liquid phase
and  the smaller size particles tending to the gas phase. At relatively high values
of polydispersity, three-phase coexistence was observed.

\appendix

\section*{Appendix A}

\renewcommand{\theequation}{A.\arabic{equation}}

Orientationally averaged Mayer function ${\bar f}_{KK}(r,\sigma_1\sigma_2)$
was fitted empirically as a sum of Yukawa-like terms
\begin{equation}
f_m(r,\sigma_i\sigma_j)=\sum_{n=1}^{N}
{\tilde A}^{(n)}_{m}\left(\sigma_i,T\right){\tilde A}^{(n)}_{m}\left(\sigma_j,T\right)
\frac{\re^{-z^{(n)}_m\left(T\right)
\left(r-\sigma_{ij}\right)}}{z^{(n)}_m\left(T\right)r}\,,
\label{I_r_sigma}
\end{equation}
where $f_1={\bar f}_{AA}$ and $f_2={\bar f}_{BB}$,
$i,j$ denotes the species of the particles,
$\sigma_{ij}=\left(\sigma_{i}+\sigma_{j}\right)/2$, $N=6$,
and ${\tilde A}^{(n)}_m\left(\sigma,T\right)={\frac{1}{ 2}}\sqrt{z_m^{(n)}(T)/\pi}A_M^{(n)}(\sigma,T)$.
The latter quantities depend on the particle size and temperature and were fitted by
``polynomially-exponential'' functions of different forms for
first and second integrals.

The fitting was performed for the following range of parameters:
$\sigma \in \left[0.85,1.2947\right]$ and $T \in \left[0.13,0.2\right]$.
The functional dependence of ${\tilde A}^{n}_m\left(\sigma_i,T\right)$ and
$z_m^{(n)}\left(T\right)$ was chosen differently for first and
second integrals.
For the first integral
\bea
{\tilde A}^{(n)}_{1}\left(\sigma,T\right)&=&a^{(1)}_{n,1}\left(T\right)
+ a^{(1)}_{n,2}\left(T\right)x +
a^{(1)}_{n,3}\left(T\right)x^2 + a^{(1)}_{n,4}\left(T\right)x^3 \nonumber\\
&&+
a^{(1)}_{n,5}\left(T\right)\exp\left[{a^{(1)}_{n,6}\left(T\right)x
+a^{(1)}_{n,7}\left(T\right)x^2 + a^{(1)}_{n,8}\left(T\right)x^6
+a^{(1)}_{n,9}\left(T\right)x^8}\right],
\eea
where $x=\left(\sigma-\sigma_{\mathrm{min}}\right)/(\Delta\sigma)$,
$\sigma_{\mathrm{min}}=0.85$, $\Delta\sigma=1.2947-\sigma_{\mathrm{min}}=0.4447$ and
$a^{(1)}_{n,1}\left(T\right),\ldots,a^{(1)}_{n,9}\left(T\right)$
are given below
\bea
a^{(1)}_{n,i}\left(T\right) &=& \re^{{b^{(1)}_{n,i,1}}y}\left(b^{(1)}_{n,i,2} +
{b^{(1)}_{n,i,3}}y + {b^{(1)}_{n,i,4}}y^2 \right), \qquad\hspace{4.5mm} \text{for} \quad
i=1,2,3,4, \nonumber \\
a^{(1)}_{n,i}\left(T\right) &=& \left(b^{(1)}_{n,i,1} + b^{(1)}_{n,i,2}y +
{b^{(1)}_{n,i,3}}y^2 + b^{(1)}_{n,i,4}y^3\right), \qquad \text{for} \quad
i=5,6,7,8,9,
\eea
where $y={T_{\mathrm{min}}}/{T}$ and $T_{\mathrm{min}}=0.13$.

The temperature dependence of $z^{(n)}_1\left( T \right)$ reads
\be
z^{(n)}_1\left( T \right) = \left(\omega^{(1)}_{n,1} + {\omega^{(1)}_{n,2}}y +
{\omega^{(1)}_{n,3}}{y^2} + {\omega^{(1)}_{n,4}}y^3 \right).
\ee

For the second integral, the functional dependence of
$A^{(2)}_{n}\left(\sigma,T\right)$ and $z^{(2)}_n\left( T \right)$
is as follows:
\bea
A^{(n)}_{2}\left(\sigma\right)&=&a^{(2)}_{n,1} + a^{(2)}_{n,2}x +
a^{(2)}_{n,3}x^2 + a^{(2)}_{n,4}x^3 +
a^{(2)}_{n,5}\exp\left[{a^{(2)}_{n,6}x+a^{(2)}_{n,7}x^2 +
a^{(2)}_{n,8}x^6+a^{(2)}_{n,9}x^8}\right], \nonumber\\
z^{(n)}_2\left( T \right) &=& \left(\omega^{(2)}_{n,1} + {\omega^{(2)}_{n,2}}y +
{\omega^{(2)}_{n,3}}{y^2} +
{\omega^{(2)}_{n,4}}\re^{{\omega^{(2)}_{n,5}}y} \right),
\eea
where
\be
a^{(2)}_{n,i} = \re^{{b^{(2)}_{n,i,1}}y}\left(b^{(2)}_{n,i,2} +
{b^{(2)}_{n,i,3}}y + b^{(2)}_{n,i,4}\re^{b^{(2)}_{n,i,5}y}\right),
\qquad \text{for} \quad i=1,2,3,4
\ee
and
\be
a^{(2)}_{n,i} = \left(b^{(2)}_{n,i,1} + b^{(2)}_{n,i,2}y +
{b^{(2)}_{n,i,3}}y^2 + b^{(2)}_{n,i,4}\re^{b^{(2)}_{n,i,5}y}\right),
\qquad \text{for} \quad i=5,6,7,8,9.
\ee
The $x$ and $y$ are the same as for the first integral.

\begin{figure}[!b]
\begin{center}
\includegraphics[clip=true,width=0.485\textwidth]{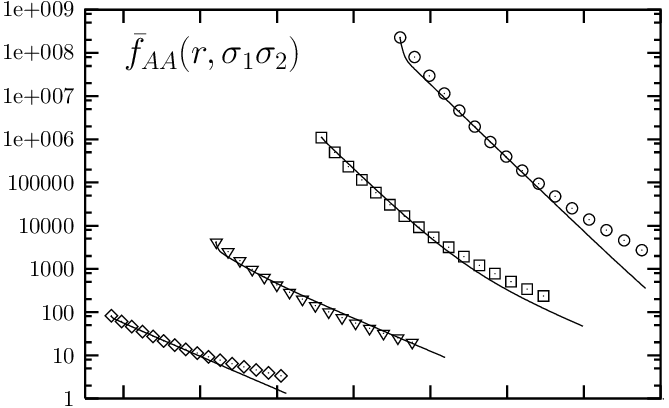}\\
\hspace{2.5mm}
\includegraphics[clip=true,width=0.475\textwidth]{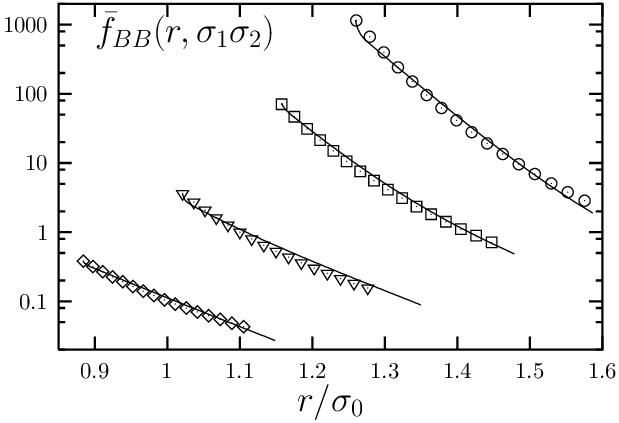}
\caption{Comparison of the exact values
of the orientation averaged Mayer functions ${\bar f}_{AA}(r,\sigma_1\sigma_2)$
(upper panel) and ${\bar f}_{BB}(r,\sigma_1\sigma_2)$ (lower panel) with their fitted
counterparts at $T^*=0.153$ and
$\sigma_1=\sigma_2=0.884\sigma_0$ (diamonds), $1.021\sigma_0$ (triangles),
$1.158\sigma_0$ (squares), $1.26\sigma_0$ (circles).\label{f13}}
\end{center}
\end{figure}

The fitting procedure consisted in finding suitable $b_{n,i,j}$
and $\omega_{n,j}$ parameters for first and second integrals by
means of differential evolution optimization algorithm~\cite{storn1997differential, price2005differential}.
The objective function in both cases was a sum of square
deviations of the area under $f_m(r,\sigma_i\sigma_j)$
as a function of $r$, and contact value
$f_m(r=\sigma_{ij},\sigma_i\sigma_j)$,
from their fitting representations
for different values of $\sigma_i$, $\sigma_j$
and $T$. The deviations in the objective function were calculated
for ten different temperature values (uniformly distributed
from $0.13$ to $0.2$) and twenty different $\sigma$ values
(also uniformly distributed from $0.85$ to $1.2947$).
For illustration purposes in figure~\ref{f13} we present a comparison of the exact values
of the orientation averaged Mayer functions ${\bar f}_{AA}(r,\sigma_1\sigma_2)$
and ${\bar f}_{BB}(r,\sigma_1\sigma_2)$ with their fitted versions at
$T^*=0.153$ and four different values for the hard-sphere size:
$\sigma_1=\sigma_2=0.884\sigma_0,\;1.021\sigma_0,\;1.158\sigma_0,\;1.26\sigma_0$.
The numerical values of the fitting parameters $b_{n,i,j}$
and $\omega_{n,j}$ can be obtained from the authors upon request.

\section*{Appendix B}
\renewcommand{\theequation}{B.\arabic{equation}}
\setcounter{equation}{0}

\begin{eqnarray}
m_l(\sigma) &=&\sigma^l,\\
m_{K,l}^{(n)}(\sigma)&=&\sigma^l\varphi\left(z_K^{(n)},\sigma\right),\\
m_{K,l}^{(n0)}(\sigma)&=&-{2\sigma^lA^{(n)}_K(\sigma)\over 1+\kappa_K^2(\sigma)}
\left[1-{2\kappa_K^2(\sigma)\over 1+\kappa_K^2(\sigma)}\right]
\left[1-\kappa_K(\sigma)\right],\\
m_{K,l}^{(nm)}(\sigma)&=&{2\sigma^lA^{(n)}_K(\sigma)A^{(m)}_K(\sigma)\over 1+\kappa_K^2(\sigma)}
\left[1-{2\kappa_K^2(\sigma)\over 1+\kappa_K^2(\sigma)}\right],\\
m_\mu(\sigma)&=&\sum_{K=A}^B\left[1+2\kappa_K^2(\sigma){1-\kappa_K(\sigma)\over 1+\kappa_K^2(\sigma)}
\right]{1-\kappa_K(\sigma)\over 1+\kappa_K^2(\sigma)}\,,
\end{eqnarray}

\begin{eqnarray}
{\delta P_{K,1}^{(n)}\over \delta F(\sigma)}&=&-{\rho\over 2z_K^{(n)}D_K^{(n)}}
\left\{\pi\sigma\left[\sigma^2+\varphi\left(z_K^{(n)},\sigma\right)\right]+2z_K^{(n)}P_{K,1}^{(n)}
{\delta D_K^{(n)}\over \delta F(\sigma)}\right\},\\
{\delta P_{K,2}^{(n)}\over \delta F(\sigma)}&=&{\rho\over 4z_K^{(n)}D_K^{(n)}}
\left\{\pi\left[\sigma^2+2\varphi\left(z_K^{(n)},\sigma\right)\right]-4z_K^{(n)}P_{K,2}^{(n)}
{\delta D_K^{(n)}\over \delta F(\sigma)}\right\},\\
{\delta P_{K,3}^{(n)}\over \delta F(\sigma)}&=&{\rho\over (2z_K^{(n)})^2D_K^{(n)}}
\left\{\pi\sigma^2\left[{2\over 3}\sigma+z_K^{(n)}\varphi\left(z^{(n)}_K,\sigma\right)\right]-
2\left(z^{(n)}_K\right)^2P_{K,3}^{(n)}{\delta D_K^{(n)}\over \delta F(\sigma)}\right\},
\end{eqnarray}

\begin{eqnarray}
{\delta D_K^{(n)}\over \delta F(\sigma)}&=&{1\over 6}\pi\sigma^3\Biggl\{m_{K,1}^{(n)}-
{3\over z_K^{(n)}}\left(m_{K,0}^{(n)}+{\frac{1}{ 2}}m_2\right)-\Delta\sigma
\left[{1\over 6}\sigma^2+\varphi\left(z_K^{(n)},\sigma\right)\right]\nonumber\\
&&-\left[{\frac{1}{ 2}}\sigma^2+\varphi\left(z_K^{(n)},\sigma\right)\right]
\left[{1\over z_K^{(n)}}\left(1+{\frac{1}{ 2}}\pi m_3\right)+{\frac{1}{ 2}}\pi m_{K,2}^{(n)}\right]\nonumber\\
&&-{1\over 4}\pi\sigma^2\varphi\left(z^{(n)}_K\right)\left(m_2+2m^{(n)}_{K,0}\right)+
{\pi\over 2}\sigma m^{(n)}_{K,1}\varphi\left(z^{(n)}_K,\sigma\right)\Biggr\},
\end{eqnarray}

\be
\varphi(z,\sigma)={1\over z}\left(1-z\sigma-\re^{z\sigma}\right),
\ee

\begin{eqnarray*}
\left[M^{(K)}_{1,1}\right]_{nm}&=&\delta_{nm}-P_{K,1}^{(m)}m_{K,1}^{(nm)}
+P_{K,3}^{(m)}m_{K,0}^{(nm)}\, ,\\
\left[M^{(K)}_{1,2}\right]_{nm}&=&           -P_{K,1}^{(m)}m_{K,0}^{(nm)}+P_{K,2}^{(m)}m_{K,1}^{(nm)}\, ,\\
\left[M^{(K)}_{2,1}\right]_{nm}&=&           -P_{K,1}^{(m)}m_{K,2}^{(nm)}+P_{K,3}^{(m)}m_{K,1}^{(nm)}\, ,\\
\left[M^{(K)}_{2,2}\right]_{nm}&=&\delta_{nm}
-P_{K,1}^{(m)}m_{K,1}^{(nm)}+P_{K,2}^{(m)}m_{K,2}^{(nm)}\, .
\end{eqnarray*}

\newpage

\ukrainianpart

\title{Фазова поведінка рідина-газ полідисперсної суміші дипольних твердих
сфер: узагальнена термодинамічна теорія збурень для асоціативного потенціалу типу центральних сил}

\author{Ю.В. Калюжний\refaddr{label1}, С.П. Глушак\refaddr{label1,label2},
П.T. Каммінгс\refaddr{label2,label3}}
\addresses{
\addr{label1} Інститут фізики конденсованих систем, Україна, 79011 Львів, вул. Свєнціцького, 1
\addr{label2} Університет Вандербільда, Теннесі, 37235 Нешвіл
\addr{label3} Інститут теорії наноматеріалів, Центр наук по нанофазних матеріалах,
Національна Лабораторія в Оук Ріджі, Оук Рідж, Теннесі, 37830}

\makeukrtitle

\begin{abstract}
\tolerance=3000%
Проведений розрахунок фазової діаграми рідина-газ полідисперсної суміші дипольних твердих
сфер з полідисперсністю як по розмірах твердих сфер, так і по величині дипольних моментів,
використовуючи узагальнення термодинамічної теорії збурення для систем з центральним характером
асоціативної взаємодії. Для того, щоб встановити зв'язок з фазовою поведінкою фероколоїдних
дисперсій, було зроблено припущення про те, що дипольний момент частинки є пропорційний до
кубу її діаметра. Ми представили та обговорили повну фазову діаграму, яка включає криві `хмари'
та `тіні', бінодалі та функції розподілу співіснуючих дочірніх фаз при різних значеннях
полідисперсності системи. У всіх випадках, які досліджувалися, полідисперсність збільшує
область фазової нестабільності та зміщує критичну точку в область вищих температур та густин.
Частинки більшого розміру завжди фракціонують у рідинну фазу, а частинки меншого розміру
віддають перевагу газовій фазі. У випадку відносно високого значення полідисперсності системи
було відмічено наявність співіснування трьох фаз.
\keywords
ТТЗ, асоційована рідина, полідисперсність, фероколоїди, фазова діаграми

\end{abstract}

\lastpage
\end{document}